\documentclass[twocolumn,secnumarabic,amssymb, nobibnotes, aps, pra]{revtex4-1}

\usepackage{braket}
\usepackage{graphicx}
\usepackage{color,soul}
\usepackage{amsmath}
\usepackage{amssymb}

\begin{document}

\title{Heralded generation of high-purity ultrashort single photons \\ in programmable temporal shapes}

\author{Vahid Ansari$^{1}$}
\email{vahid.ansari@uni-paderborn.de}
\author{Emanuele Roccia$^{2}$}
\author{Matteo Santandrea$^{1}$}
\author{Mahnaz Doostdar$^{1}$}
\author{Christof Eigner$^{1}$}
\author{Laura Padberg$^{1}$}
\author{Ilaria Gianani$^{2}$}
\author{Marco Sbroscia$^{2}$}
\author{John M. Donohue$^{1}$}
\author{Luca Mancino$^{2}$}
\author{Marco Barbieri$^{2}$}
\author{Christine Silberhorn$^{1}$}

\affiliation{$^1$Integrated Quantum Optics, Paderborn University, Warburger Strasse 100, 33098 Paderborn, Germany}
\affiliation{$^2$Dipartimento di Scienze, Universit`a degli Studi Roma Tre, Via della Vasca Navale 84, 00146, Rome, Italy}


\begin{abstract}
We experimentally demonstrate a source of nearly pure single photons in arbitrary temporal shapes heralded from a parametric down-conversion (PDC) source at telecom wavelengths. The technology is enabled by the tailored dispersion of in-house fabricated waveguides with shaped pump pulses to directly generate the PDC photons in on-demand temporal shapes. We generate PDC photons in Hermite-Gauss and frequency-binned modes and confirm a minimum purity of 0.81, even for complex temporal shapes.
\end{abstract}

\maketitle

Preparing single photons in pure and controlled spectral-temporal modes is a key requirement for quantum photonic technologies.
Diverse applications including quantum-enhanced metrology \cite{boto2000quantum,Treps2002}, quantum computation \cite{menicucci2008one, Rahimi-Keshari2016}, and quantum encryption \cite{Jennewein2000, Nunn2013, Zhong2015} rely on high-contrast interference through stable sources of pure single photons. In addition, widely customisable and precisely controllable temporal-mode shaping is necessary to ensure mode matching between individual sources \cite{Mosley2008b}, facilitate coupling between nodes in a quantum network \cite{kielpinski}, and enable temporal-mode based quantum communication \cite{Brecht2015} and source mupliplexing \cite{Kaneda2015,Francis-Jones2016}, among other applications. Furthermore, sources with high brightness are essential for scalable performance, and spatially single-mode behaviour is necessary for coupling to optical fibre networks and integrated waveguide devices.

Sources based on parametric downconversion (PDC) have granted a simple solution to heralded single-photon generation for decades, but have not yet satisfied all of the above requirements simultaneously. Most PDC sources generate photons with strong spectral correlations which is undesirable for heralded single-photon sources. However, it is possible to minimise the spectral correlation in crystals offering specific dispersion properties along with an adapted pump bandwidth \cite{Grice2001, Walton2003,Kuzucu2005,Mosley2008b,Evans2010,Eckstein2011b,Jin2013,Harder2013,Kaneda2016,Spring2017}. This specific dispersion property is linked to the group velocities of the pump and the PDC photons and can be summarised in two categories: matching the group velocity of the pump photon with one of the PDC photons \cite{Mosley2008b,Kaneda2016}, or having the group velocity of the pump between the two PDC photons \cite{Kuzucu2005,Eckstein2011b,Jin2013,Harder2013}.

\begin{figure}[tbp]
\centering
\includegraphics[width=1\linewidth]{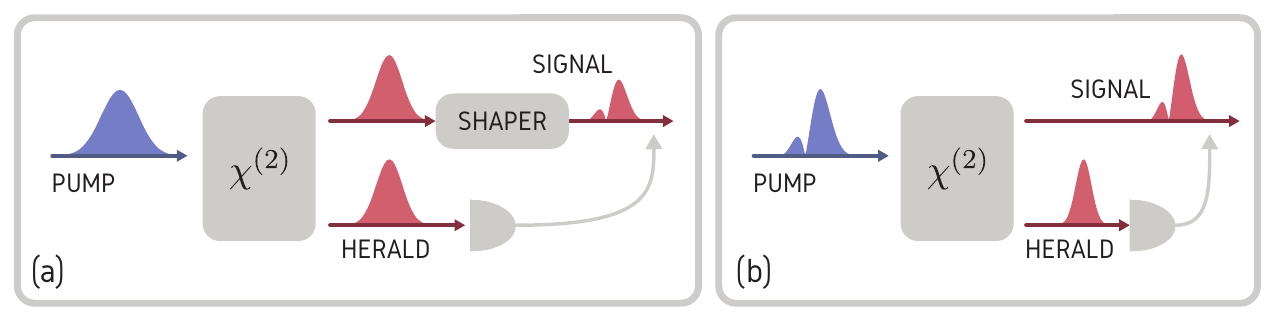}
\caption{Heralded source of temporally shaped single-photons. (a) The desired temporal mode can be carved out of PDC photons after the generation, which inevitably reduces the heralding efficiency. (b) With an appropriately designed pump field and group-velocity engineered nonlinear medium, the PDC photons are emitted directly in a desired temporal shape. In both scenarios the purity of heralded single photon rely on the separability of the PDC state in terms of signal and herald fields.}
\label{fig:concept}
\end{figure}

On the other hand, efficient temporal-mode shaping of the PDC photons is more challenging. Existing methods to create a broadband single photon in an arbitrary temporal mode rely on carving out the desired mode from the original wavepacket as depicted in Fig. \ref{fig:concept}(a), which can be accurately achieved by temporal or spectral modulation of the photon \cite{Bellini2003, PeEr2005, Baek2008, Lukens2014, Bernhard2013}. This method, however, necessarily introduces loss and leads to a reduced rate of prepared photons \cite{URen2003} and a low pair-symmetric heralding efficiencies \cite{Meyer-Scott2017}; this poses a practical limit for many experiments such as device-independent quantum cryptography \cite{acin2007device,Lucamarini2012} and optical quantum computing \cite{Varnava2008,Gong2010,Jennewein2011}. Temporal manipulation is also possible with shaped-pulse mediated nonlinear interactions \cite{kielpinski,donohue2016spectrally, Matsuda2016, Allgaier2017, Averchenko2017}, Raman interfaces \cite{fisher2016frequency}, or ultrafast electro-optic modulation \cite{karpinski2016bandwidth}, but these methods are experimentally challenging to implement without prohibitive loss. To minimise the potential for photon loss, a source which generates heralded photons in a customisable and pure spectral state is highly desirable.

In this letter, we take a novel approach to directly create PDC photons with tailored temporal-modes. Through group-velocity matching two of the interacting fields in the PDC process, we generate heralded photons which inherit the temporal shape of the pump pulse, as sketched in Fig. \ref{fig:concept}(b). We show through joint spectral measurements and second-order photon number correlations that the photons are generated in a highly pure state. We explicitly demonstrate the versatility of our source design by generating photons with customised temporal shapes, such as broadband Hermite-Gaussian temporal modes and narrow frequency bins. Our source is based on the in-house fabricated unpoled KTP waveguides and emits in the near-infrared telecommunications regime, making it a prime candidate for use in long-distance quantum protocols and fibre-based networks. Our result bridges an important gap in quantum state engineering of time-frequency modes, and enables a range of quantum photonic applications that require temporal-mode matching.

\section{Theory}
\label{sec:theory}

\begin{figure}[tbp]
\centering
\includegraphics[width=1\linewidth]{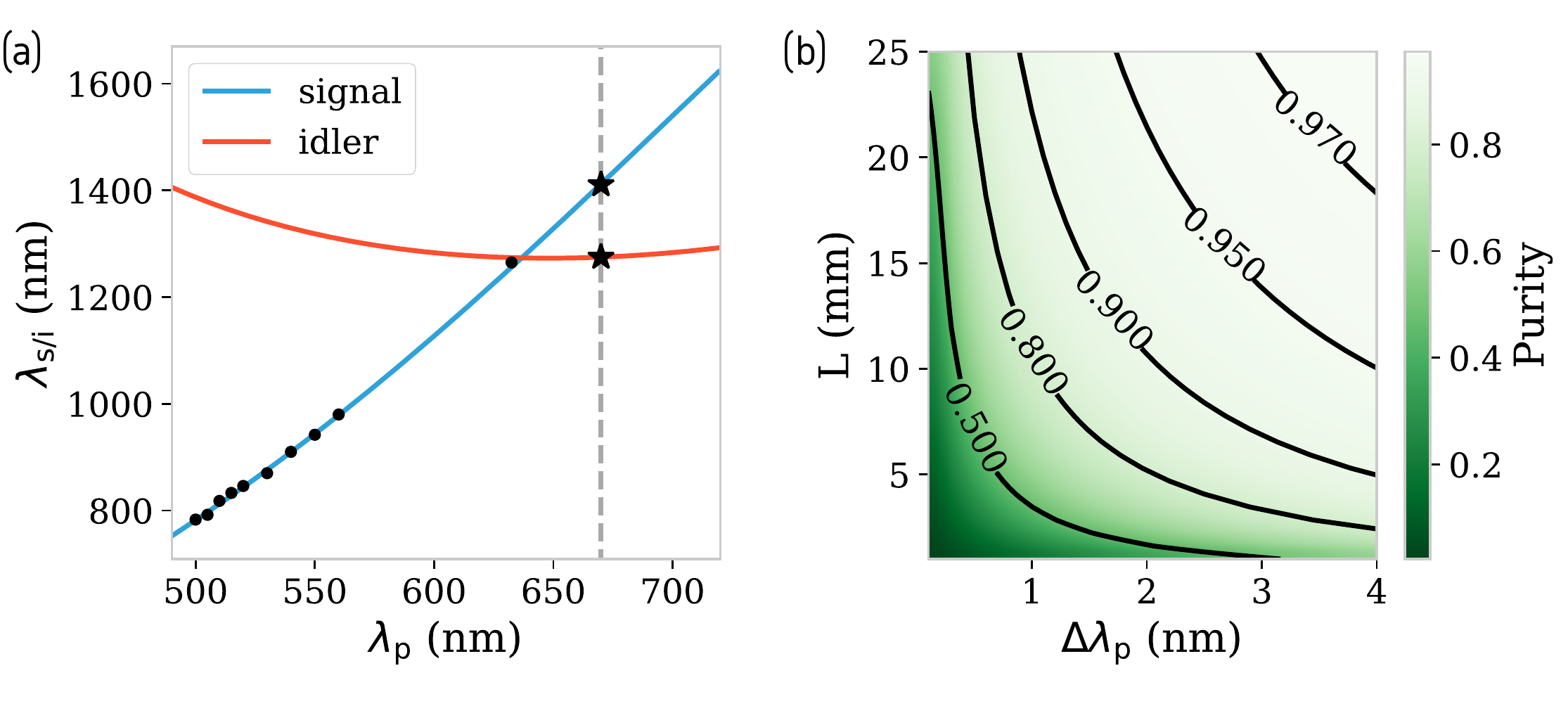}
\caption{(a) Birefringent phasematched type-II PDC processes in the KTP waveguide versus pump wavelength. The pump and idler photons are TE polarised and the signal photon is TM polarised. The dots correspond to experimentally measured PDC photons with a tunable pulsed pump laser and a single-photon sensitive spectrometer. The data point at the degeneracy point, however, is measured by means of second harmonic generation with a pulsed pump at the central wavelength of 1275 nm. The error bars are smaller than the markers. To generate a single-mode JSA we use the AGVM condition which holds for a pump wavelength of 670 nm (indicated with the vertical dashed line) and a signal and idler wavelengths of 1411 nm and 1276 nm, respectively (star markers). (b) Theoretical spectral purity of the JSA for different pump bandwidths $\Delta\lambda_\mathrm{p}$ and crystal lengths $L$.}
\label{fig:purity}
\end{figure}

The PDC process in the waveguides happens when the energy conservation $\omega_{\mathrm{p}}=\omega_{\mathrm{s}}+\omega_{\mathrm{i}}$ and the momentum conservation $k_{\mathrm{p}}=k_{\mathrm{s}}+k_{\mathrm{i}}$ between the three fields --- pump, signal, and idler --- are satisfied. The momentum conservation with waveguided collinear propagation is typically achieved with quasi-phasematching through periodic poling. 
In our case, we consider a type-II PDC process where momentum conservation is enabled by birefringent phasematching without a need for periodic poling. The waveguided structure we consider is made with a z-cut KTP substrate and waveguides fabricated along the x-axis of the crystal with rubidium ion exchange. To calculate the effective refractive indices of the optical fields inside of the Rb:KTP waveguide, we use a commercial finite-element mode solver along with a model for the refractive index profile provided in \cite{Callahan2014}. Theoretically calculated phasematched type-II processes for different pump wavelengths are plotted in Fig. \ref{fig:purity}(a) as solid lines, where we also experimentally verified our model (see the caption).

The Hamiltonian of type-II PDC process is
\begin{equation}
\hat{H}_\mathrm{PDC} \propto \iint f(\omega_\mathrm{s},\omega_\mathrm{i}) \hat a_{\mathrm{TM}}^\dagger(\omega_\mathrm{s}) \hat a_{\mathrm{TE}}^\dagger(\omega_\mathrm{i}) d\omega_\mathrm{s} d\omega_\mathrm{i}  + \mathrm{h.c.},
\label{eq:hamiltonian}
\end{equation}
where $a^\dag(\omega)$ is the standard creation operator at frequency $\omega$. The joint spectral amplitude (JSA) function
\begin{equation}
f(\omega_\mathrm{s},\omega_\mathrm{i}) = \alpha(\omega_\mathrm{s} + \omega_\mathrm{i}) \phi(\omega_\mathrm{s}, \omega_\mathrm{i}),
\label{eq:jsa}
\end{equation}
describes the spectral-temporal properties of the PDC state, where $\alpha(\omega_\mathrm{s} + \omega_\mathrm{i})$ is the ultrashort pump amplitude function and $\phi(\omega_\mathrm{s},\omega_\mathrm{i})$ is the phasematching function expressing the momentum conservation between the fields in the waveguide. 
Due to energy conservation, PDC sources typically exhibit spectral correlations. However, many applications benefit from spectrally pure single-photon states with separable JSAs of the form
\begin{equation}
f(\omega_\mathrm{s},\omega_\mathrm{i}) \approx g(\omega_\mathrm{s}) h(\omega_\mathrm{i}),
\label{eq:sm_jsa}
\end{equation}
which can be achieved by dispersion engineering.

The strength of spectral correlations in the PDC state can be quantified by a Schmidt decomposition of the JSA function\cite{Law2000, Eberly2006}. This defines the Schmidt number $K$ as the effective number of temporal-modes in the state. An experimentally accessible method to measure the Schmidt number and the purity $\mathcal{P}$ of the PDC photons is by means of second-order correlation function $g^{(2)}(\tau=0)$ of unheralded signal or idler photons as \cite{Christ2011}
\begin{equation}
\mathcal{P} = \frac{1}{K} = g^{(2)}(0)-1.
\label{eq:purity}
\end{equation}
In the case of spectrally pure PDC state with $K=1$, the partial trace of the PDC state exhibits thermal photon number statistics corresponding to $g^{(2)}(0)=2$.
With a multimode state, we would measure a convolution of all the different thermal photon statistics, since the detector cannot discriminate each mode, which results in a Poissonian photon-number distribution and a $g^{(2)}(0)$ that approaches 1 \cite{Mauerer2009}.

\begin{figure}[t]
\centering
\includegraphics[width=1\linewidth]{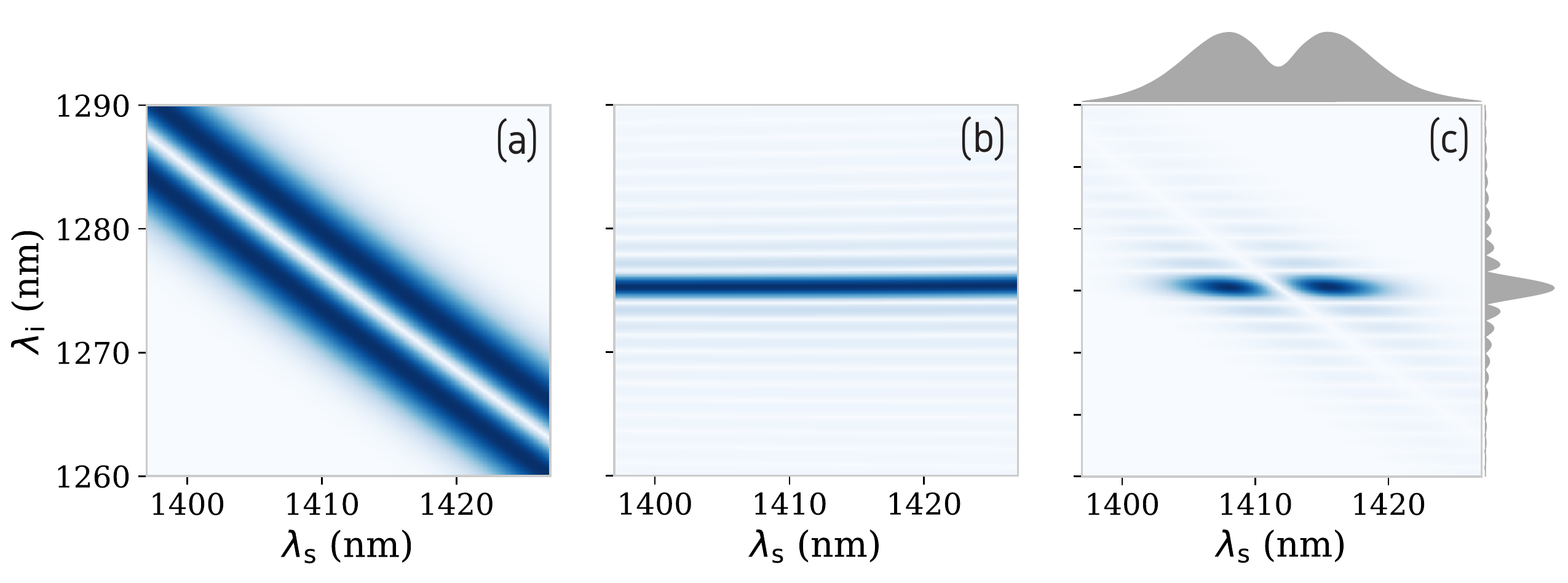}
\caption{(a) The absolute value of the pump spectrum $|\alpha(\omega_\mathrm{p} = \omega_\mathrm{s} + \omega_\mathrm{i})|$ with the first-order Hermite-Gaussian profile with FWHM of 2 nm. (b) Phasematching function $|\phi(\omega_\mathrm{s}, \omega_\mathrm{i})|$ of a KTP waveguide with a length of 16 mm. (c) Theoretical joint spectral amplitude $|f(\omega_\mathrm{s},\omega_\mathrm{i})|$ of the PDC state and its marginal distributions. The modelled JSA shows a Schmidt number of $K=1.087$ and a spectral purity of $\mathcal{P}=0.919$. All functions are plotted against wavelengths (instead of angular frequencies $\omega_j$) to provide a convenient comparison with the experimental data.
}
\label{fig:jsa}
\end{figure}

To realise a single-mode JSA we exploit a phasematching with matched group-velocities of pump and signal fields, known as the asymmetric group-velocity matching (AGVM) condition \cite{UextquoterightRen2005}. This condition holds for a pump wavelength of 670 nm (TE polarised) and a signal wavelength of 1411 nm (TM polarised), which are marked with stars in Fig. \ref{fig:purity}(a). In order to find the experimental settings for an optimum spectral purity, we calculate $\mathcal{P}$ for different pump pulse bandwidths and crystal lengths as plotted in Fig. \ref{fig:purity}(b). In our experimental implementation, we use the AGVM condition with a pump spectral FWHM of 2 nm and a crystal length of 16 mm. This configuration leads to a nearly single-mode JSA as plotted in Fig. \ref{fig:jsa}, where we plot the JSA function with the pump in the first-order Hermite-Gauss mode. From the JSA and its marginal distributions in Fig. \ref{fig:jsa}(c), it is clear that the phasematching function is mapped onto the idler photon, while the spectral profile of the pump is imparted into the signal photon. A similar AGVM condition can be also achieved in a periodically poled bulk KTP but at wavelengths outside of the telecom bands \cite{Kaneda2016}. Additionally, the waveguided structure, in comparison to bulk, accommodates a longer interaction length and a stronger field confinement, allowing for higher parametric gains and narrower phasematching functions. Note that a narrow phasematching is crucial for high fidelity shaping of the signal photon.

\section{Experiment}
\begin{figure}[tbp]
\centering
\includegraphics[width=1\linewidth]{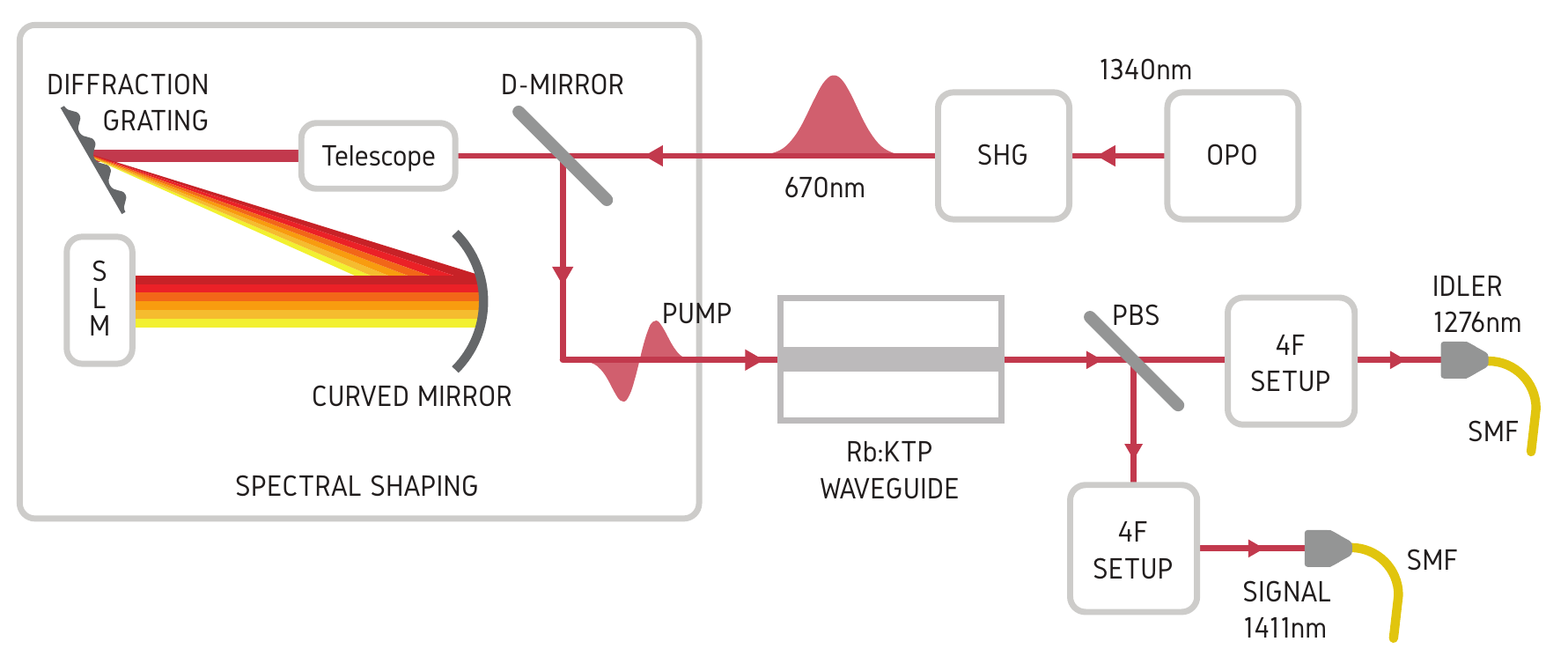}
\caption{Experimental setup. To prepare ultrashort pump pulses at 670 nm we take second harmonics (SHG) of an optical parametric oscillator (OPO). In the spectral shaping setup, we use a reflective spatial light modulator (SLM) in a folded 4f-setup to shape the spectral amplitude and phase of the pump field. A telescope is used to match the size of each frequency component with SLM's pixels to get an optimum resolution. The SLM reflects the beam at a slightly different angle which displaces the reflected beam vertically and allows us to collect the reflected beam with a d-shaped mirror. The generated PDC photons are separated on a broadband polarising beamsplitter (PBS). We use 4f-setups for both PDC photons to filter the undesirable background. Finally each beam is coupled into single-mode fibres (SMF) for telecom wavelengths.}
\label{fig:setup}
\end{figure}

The outline of the experimental setup is given in Fig. \ref{fig:setup}. To prepare the pump of the PDC process, we take ultrashort pulses at the central wavelength of 670 nm (from a frequency doubled optical parametric oscillator) (Coherent Chameleon OPO with APE HarmoniXX) and use a pulse shaper to carve out the desired temporal modes. The pulse shaper is a folded 4f-setup consist of a magnifying telescope, a holographic diffraction grating with 2000 lines per mm (Spectrogon), a cylindrical silver mirror and a two-dimensional reflective liquid crystal on silicon spatial light modulator (Hamamatsu X10468-07 LCoS-SLM) \cite{Vaughan2005, Frumker2007}. This 4f-setup has a spectral resolution of 35 pm which along with the initial 6 nm bandwidth of our laser system, allows us to accurately prepare e.g. Hermite-Gaussian pulses of up to fourth order with 2 nm of FWHM for the Gaussian profile. We use spectral interferometry to characterise the performance of the pulse shaper and ensure a dispersion-free alignment \cite{Takeda1982}.

The heart of the experiment is a 16 mm long in-house built Rb:KTP waveguide with a nominal width of 3 $\mu$m and depth of 5$\mu$m, designed to be spatially single-mode over the whole telecom range for both TE and TM polarisations. 
The sample is produced in a two-step process. At first, KTP sample with dimensions of about 20~$\times$~6~$\times$~1~mm\textsuperscript{3} is immersed in a pure KNO\textsubscript{3} melt to homogenize the sample composition. In a second step, after a titanium mask is patterned on the +c face of the sample using standard photolithography to define waveguide structures with different widths, the sample is immersed in a RbNO\textsubscript{3}/KNO\textsubscript{3}/Ba(NO\textsubscript{3})\textsubscript{2} melt, where K\textsuperscript{+} ions in the crystal are exchanged with Rb\textsuperscript{+} ions present in the melt. After removing the Ti mask, the sample facets are polished to provide a smooth surface suitable for free space coupling and fibre pigtailing.

To couple the laser to the waveguide, we use a aspheric lens with a focal length of 8 mm. Using the Fabry-Perot interferometric method \cite{Regener1985} we measure internal waveguide average losses of 0.85~dB/cm (with a minimum of 0.66~dB/cm and a maximum of 1.15~dB/cm) and 0.67~dB/cm (with a minimum of 0.53 dB/cm and a maximum of 0.78~dB/cm) at 1550 nm for TE and TM polarisations, respectively. To estimate the maximum achievable coupling efficiency of the waveguide mode into the standard single-mode telecom fibre (SMF-28) we use bright lasers matched with central frequencies of the PDC photons and measure coupling efficiencies of 0.65 and 0.60 for idler (TE) and signal (TM) photons, respectively. This is due to asymmetry of the waveguide mode which can be designed to be more symmetric at any chosen wavelength by modifying the fabrication parameters e.g. the diffusion depth. The waveguide used in this work has a more symmetric mode profile at 1550 nm where we measure a coupling efficiency of more than 0.80.

\subsection{Heralding efficiency and brightness}
To spectrally filter the pump field and the parasitic background noise, we use folded 4f-setups in each arm, aligned around the central frequencies of PDC photons. The total transmissions of 4f-setups are 0.26 and 0.30 for signal and idler photons, respectively, owing principally to a low diffraction efficiency of the diffraction gratings. The PDC photons are then detected with fibre coupled superconducting nanowire single photon detectors (SNSPD) (Photon Spot) with system detection efficiencies of 0.41 and 0.55 at 1276 nm and 1411 nm, respectively. With this configuration we measure Klyshko efficiencies (coincidences over the single counts of each arm \cite{Klyshko1980}) of (8$\pm$0.01)\% and (5$\pm$0.02)\% for signal and idler photons, respectively. A normalisation over the transmission of 4f-setups (which can be replaced with bandpass filters with very high transmissivities) and detection efficiencies suggests that these Klyshko efficiencies can be improved to around 56\% and 40\% for signal and idler photons, respectively. These efficiencies can be further improved by using anti-reflective coatings on the KTP waveguide facet and the fibres.

\begin{figure}[tbp]
\centering
\includegraphics[width=.6\linewidth]{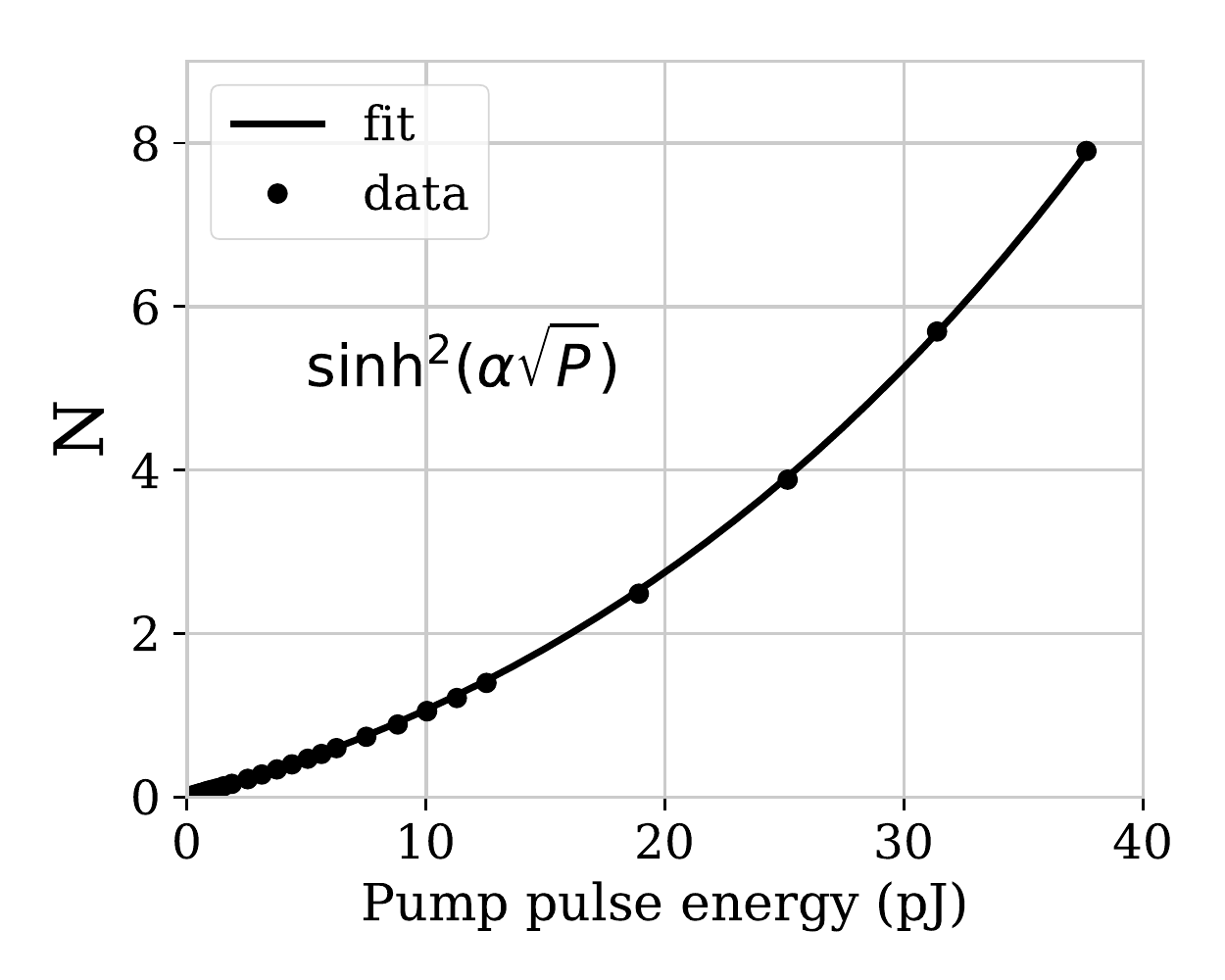}
\caption{Mean photon number N of one PDC arm versus pump pulse energy. The pump pulse energy is measured after the waveguide to account for the incoupling loss. The only fit parameter used for the fitting function $\mathrm{sinh}^2(\alpha\sqrt{P})$ is $\alpha=0.28$.}
\label{fig:meanN}
\end{figure}

To benchmark the brightness of the source, in in Fig. \ref{fig:meanN} we plot the generated mean photon number versus pump pulse energy. For this measurement, we calibrate the detected counts using the Klyshko method \cite{Klyshko1980}. When we drive the waveguide with a pump pulse energy of 37.5 pJ, the source generates states with a mean photon number of about 8. The mean photon number follows the expected curve for a single-mode source at relatively low pump powers. The comparably high brightness of the source is due to long waveguide length, the use of birefringent phasematching, and the relatively single-mode character of the source.

\begin{figure}[tbp]
\centering
\includegraphics[width=1\linewidth]{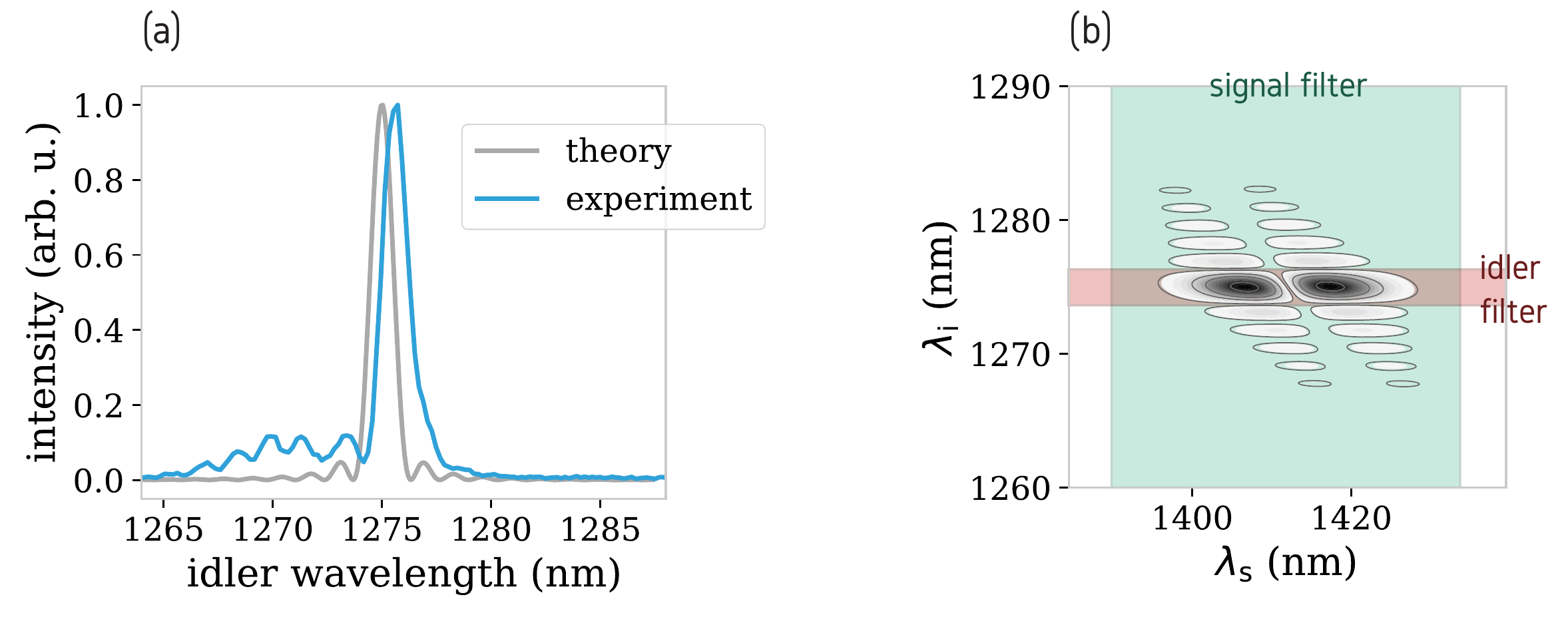}
\caption{(a) Theoretical and experimentally measured phasematching functions phasematching function $|\phi(\omega_\mathrm{s}, \omega_\mathrm{i})|$. (b) A contour plot of the theoretical JSA function and the bandpass filters to remove the phasematching sidelobes. The widths of filters in signal and idler arms are 45 nm and 3 nm, respectively. Without any spectral filtering, the JSA features a Schmidt number $K=1.12$ which increases to $K=1.03$ when the idler filter is applied. Due to the specific distribution of JSA, filtering the signal photons cannot remove the phasematching sidelobes.}
\label{fig:filter}
\end{figure}

\subsection{Spectral characterisation}
To measure the spectrum of the idler photon, we use the 4f-setup in the monochromator configuration, with a spectral resolution of 0.2 nm. With the AGVM condition, as can be seen in Fig. \ref{fig:jsa}(c), the spectrum of the idler photon echoes the phasematching function $\phi(\omega_\mathrm{s}, \omega_\mathrm{i})$. The measured spectrum of the idler photon and its theoretical counterpart are plotted in Fig. \ref{fig:filter}(a). 
The discrepancy between experiment and theory can be explained by considering the waveguide inhomogeneities \cite{Cao1991, Helmfrid1993, Bortz1994, Cristiani2001}. Inhomogeneity of the waveguide channel, e.g. non-uniform width or depth, can change the effective refractive index along the propagation direction and consequently distort the phasematching function.
This can be understood by regarding the inhomogeneous waveguide as many short homogeneous segments with different phasematching conditions $\Delta k(\omega_\mathrm{s},\omega_\mathrm{i}) = k_\mathrm{p}(\omega_\mathrm{s}+\omega_\mathrm{i}) - k_\mathrm{s}(\omega_\mathrm{s}) - k_\mathrm{i}(\omega_\mathrm{i})$. The overall phasematching distribution then would be a sum of all of these segments, which is a coherent mixture of many sinc-shaped functions with different widths and central frequencies. This depends on the exact form of inhomogeneities, but in general these inhomogeneities effectively broaden the phasematching function. A known solution to this is to design the waveguide geometry insensitive to these inhomogeneities (known as noncritical phase-matching) \cite{Lim1990, Bortz1994}. We are currently conducting a comprehensive study of this matter.

The broadened phasematching with asymmetric side-lobes, seen in Fig. \ref{fig:filter}(a), diminishes the spectral purity and the fidelity of single photon shaping, hence we use spectral filtering (with a width of 3 nm) on the idler photons to remove the side-lobes, as shown in Fig. \ref{fig:filter}(b). 
From the distribution of the JSA, it is evident that removing the phasematching sidelobes is only possible by filtering the idler photons. 
This filtering, however, result in an imbalanced Schmidt number $K$ between the signal and idler photons \cite{Laiho2011}. Nonetheless, since we are interested in heralding signal photons upon detection of idler photons, the heralded photons inherent the lowest Schmidt number $K$, which consequently yields a high purity.

\begin{figure}[tp]
\centering
\includegraphics[width=1\linewidth]{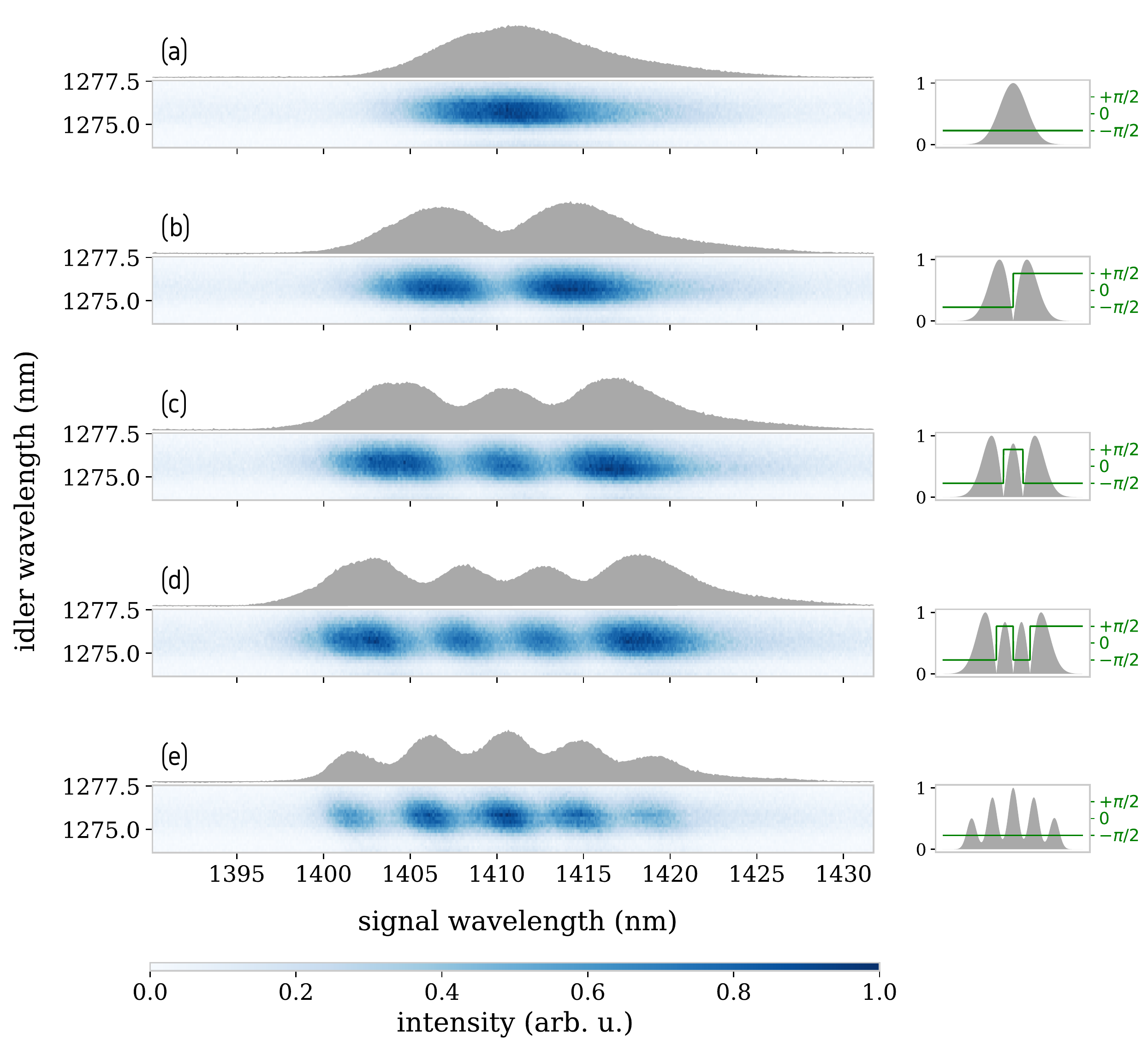}
\caption{A few examples of the measured joint spectral intensities (JSIs), with the marginal spectral distribution of signal photon above in grey. The pump mode for each JSI is shown on the right side, where the grey shaded area is the spectral amplitude and the green line is the spectral phase. The pump modes are as the following: (a) Gaussian, (b) 1st-order Hermite-Gaussian, (c) 2nd-order Hermite-Gaussian, (d) 3rd-order Hermite-Gaussian, (e) frequency bins, with Schmidt numbers: $K_a=1.01$, $K_b=1.01$, $K_c=1.02$, $K_d=1.02$, $K_e=1.02$.}
\label{fig:jsis}
\end{figure}

To measure joint spectral intensity (JSI) distribution $|f(\omega_\mathrm{s}, \omega_\mathrm{i})|^2$, we combine the monochromator in the idler arm with a time-of-flight spectrometer in the signal arm \cite{Avenhaus2009a}. In the time-of-flight spectrometer we use a highly dispersive fibre to map the spectrum into the temporal profile which can be resolved directly in time on SNSPDs. We use a 4.3 km long fibre with a total dispersion of 0.3 ns/nm which, alongside with 70 ps timing resolution of SNSPD, constitute a spectrometer with a resolution of about 0.2 nm. The measured JSIs with the pump field in the first four Hermite-Gauss modes and five frequency bins are plotted in Fig. \ref{fig:jsis}. The Schmidt number inferred from these JSIs shows a spectral purity of more than 0.98, which provides an upper bound on the spectral purity.

\begin{figure}[tbp]
\centering
\includegraphics[width=1\linewidth]{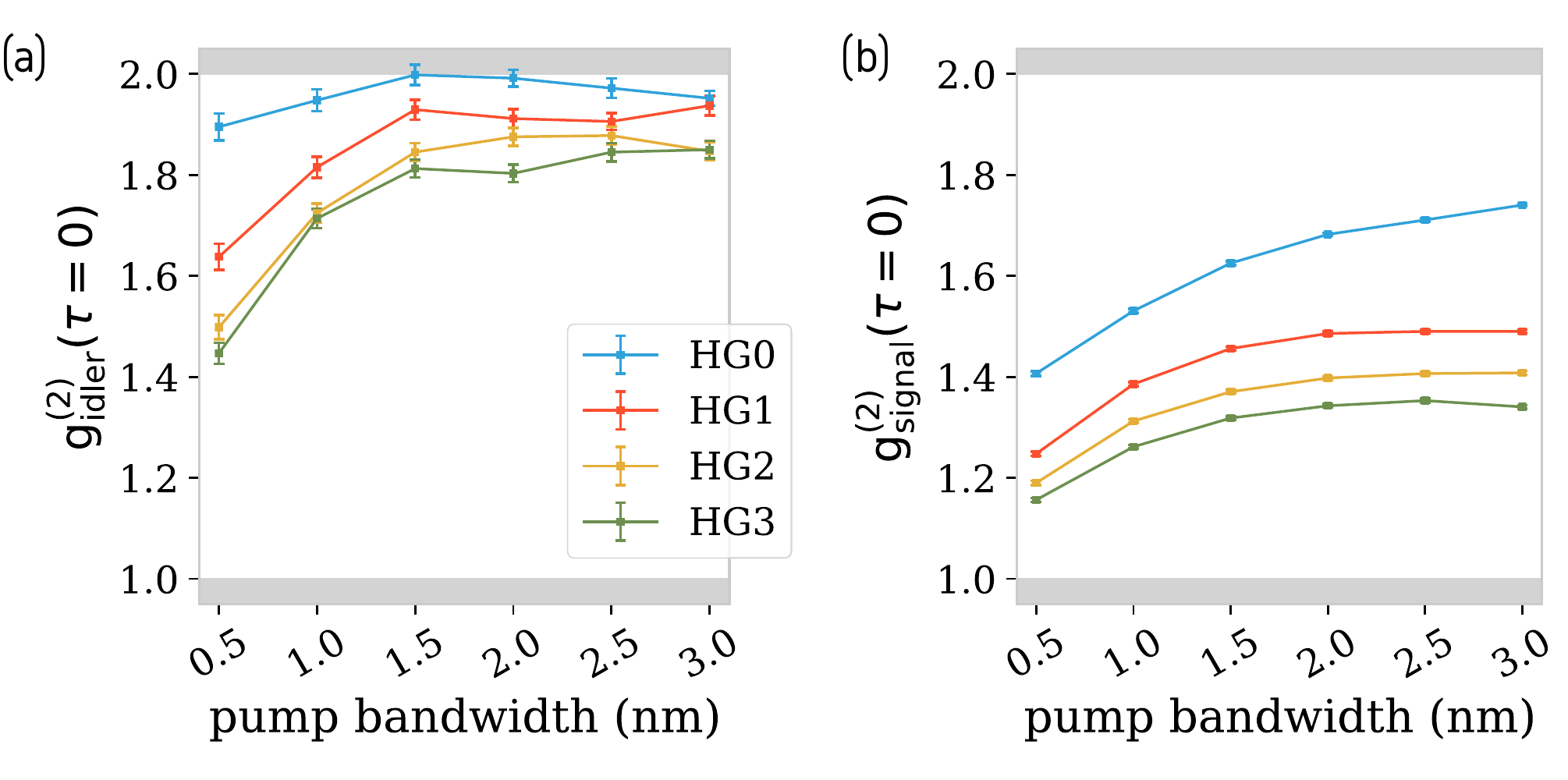}
\caption{The second-order correlation measurements of the idler and signal photons with the pump set to different bandwidths and different orders of Hermite-Gaussian modes. The error bars for the $g^{(2)}_{\mathrm{signal}}(\tau=0)$ are smaller than the markers.}
\label{fig:g2}
\end{figure}

\subsection{Purity and second-order correlation function}
As discussed in the theory section, spectral correlations between PDC photons leads to impurity of the heralded single photons. Although the JSI measurement provide important information about the spectral correlation of the PDC photons, it is blind to the spectral phase of the photons and is also limited by the resolution of the spectrometers. A better measure of any underlying correlations of the PDC photons is the second-order correlation function $g^{(2)}(0)$ of signal or idler photons, as measured with a 50/50 fibre coupler \cite{Tapster1998,Christ2011}. The $g^{(2)}(0)$ measurement probes the photon number statistics of unheralded beams (signal or idler) and can discriminate between a single-mode PDC state with $g^{(2)}(0)=2$ and a highly multimode state with $g^{(2)}(0)=1$. In Fig. \ref{fig:g2} we plot the $g^{(2)}(0)$ of the both PDC photons with the pump pulse in different orders of Hermite-Gauss modes and bandwidths ranging from 0.5 nm to 3 nm. With a narrow pump bandwidth, energy correlations remain in the PDC state which are exhibited in lower $g^{(2)}(0)$ values. For the idler photon, we spectrally filter the asymmetric phasematching side-lobes (see Fig. \ref{fig:filter}), and we achieve the highest $g^{(2)}(0)$ of $1.99\pm0.02$ with a 1.5 nm broad Gaussian pump pulse, which reduces to $1.93\pm0.02$, $1.85\pm0.02$, and $1.81\pm0.02$ for the first, second, and third order Hermite-Gauss modes, respectively. The corresponding purities can be calculated through Eq. (\ref{eq:purity}). This reduction in the $g^{(2)}(0)$ value is also expected from theory. With increasing order of Hermite-Gauss modes, these function feature more complex structures spanned over a broader frequency range which inevitably increases the frequency anti-correlations between signal and idler (see Fig. \ref{fig:jsa}(a)). Despite this, it is possible to achieve a high purity with an appropriately designed crystal length.

The $g^{(2)}(0)$ of signal photons, plotted in Fig. \ref{fig:g2}(b), is considerably lower. This is due to the presence of the phasematching side-lobes which cannot be simply filtered for the signal photons (see Fig. \ref{fig:filter}(b)). While the signal photons are themselves less pure, the high $g^{(2)}(0)$ of the idler indicates that the shaped signal photons are highly pure when heralded by an idler detection. However, this purity comes at a cost of heralding efficiency. Enhancing the waveguide fabrication technology or using methods such as noncritical phasematching \cite{Lim1990} or aperiodic poling \cite{Dosseva2016, Graffitti2017} may be able to eliminate these unwanted spectral features to produce filter-free heralded photons with high purities and arbitrary temporal shapes.

\section{Conclusion}
We have shown that heralded single photons can be generated in arbitrary temporal modes using pulse shaping and KTP waveguides with an optimised dispersion. Through joint spectral intensity measurements, we have verified that the spectral shape of the pump pulse is faithfully imparted onto the signal photon. Second-order photon number correlations measurements show that the heralded photon state is highly pure and suitable for use in quantum networks. Our integrated source is based on birefringent phasematching and emits within the telecommunications band. Future work will focus on adapting the source with periodic poling to optimise the emission wavelengths for available photon detectors and customise the joint spectral amplitude to eliminate the need for filtering. With these optimisations, the source presented here will prove to be a vital component in applications such as temporal-mode based quantum communication and mode matching between quantum interfaces.

\section{Funding}
Gottfried Wilhelm Leibniz-Preis; European Union's Horizon 2020 research and innovation programme under grant agreement No 665148. 

\section{Acknowledgments}
E.R. thanks the integrated quantum optics group for their hospitality during the completion of this work. We thank Fabio Sciarrino for fruitful discussion.

\end{document}